\documentstyle[twoside,fleqn,espcrc2]{article}
\input psfig.tex

\newcommand{\AmS}{{\protect\the\textfont2
  A\kern-.1667em\lower.5ex\hbox{M}\kern-.125emS}}

\newcommand{\fa}{f_{\rm a}}
\newcommand{\ma}{m_{\rm a}}

\title{Spectrum of radiation from axion strings}

\author{R.A. Battye and E.P.S. Shellard \address{Department of Applied Mathematics and Theoretical Physics, University of Cambridge, \\ Silver Street, Cambridge, CB3 9EW, U.K.}}
       
\begin{document}

\begin{abstract}
In the wide variety of axion cosmologies in which axion strings form, their 
radiative decay is the dominant mechanism for the production of axions, imposing a tight constraint on the axion mass. 
Here, we focus on the mechanism by which axions are produced in this scenario and, in particular, the key issue of  
the axion spectrum emitted by an evolving network of strings.
\end{abstract}

\maketitle

\section{INTRODUCTION}

The spectrum of radiation emitted by global cosmic strings is of crucial importance to viability of axionic cold dark matter scenarios.
There are a broad variety of scenarios in which a network of cosmic
strings are formed in the early universe, irrespective of whether 
or not in an inflationary context the reheat temperature of the
universe is above the Peccei-Quinn scale $\fa$ (see
ref.~\cite{BatShe98}).  The axion string network evolves quickly to a scaling regime in which there
are a fixed number of long strings stretching across each horizon volume.  
This overall energy loss is maintained by the production of loops 
and subsequent emission of massless axion radiation. Around the QCD phase transition, `soft' 
instanton effects break the residual discrete symmetry and the axions become massive. This process also produces 
domain walls connected to the strings; a hybrid system which is unstable and collapses within one Hubble time, if 
there is a unique CP conserving vacuum ($N$$=$$1$). 

The important issue when considering the contribution of axions to the mass density of the universe is 
the effect of high frequency modes and the literature on this subject has traditionally split into two 
unequal camps. The original work~\cite{Dav86} assumed that the spectrum was dominated by modes
close to the fundamental frequency of the perturbation, a contention which has been broadly supported by 
numerical simulations of axion radiation from an oscillating string~\cite{DavShe89}, plus a detailed 
comparison between more extensive numerical simulations and analytic predictions treating the strings as 
line-like objects within the framework of the Kalb-Ramond action~\cite{BatShe94a}. Under these assumptions 
one would see a perturbed string oscillate, losing energy all the time, with the maximum amplitude of 
oscillation being gradually damped. If one assumes that the dominant mechanism for the production of 
axions is through the collapse of string loops~\cite{BatShe94b,BatShe97}, then the
prediction obtained by assuming $\Omega_{\rm a}\sim 1$ is that the
axion mass $\ma \sim 100\mu$eV, with possible uncertainties, most of
which are common to all calculations of the cosmological axion 
density; these are about a factor of 5 in each direction (see
ref.\cite{BatShe98}).

This widely accepted viewpoint has been repeatedly questioned by Sikivie and 
collaborators~\cite{HarSik87,HagSik91,HagSik98}, who maintain that the power emitted in the 
$n$th mode falls-off like $P(n)\propto n^{-1}$ for large $n$, that is, a flat spectrum per 
logarithmic frequency interval. A physical consequence of this prediction is that a perturbed 
string would straighten itself in one oscillation, that is, it would be critically damped. This 
divergent spectrum leads to a prediction which systematically differs from the standard prediction 
by up to two orders of magnitude $\ma\sim 1\mu{\rm eV}$ with similar levels of uncertainty.

These two very different predictions for the axion mass can be thought of as being due to one 
universal difference between the predicted spectra. If one considers a spectrum  of radiation 
$P(n)\propto n^{-q}$, then the calculation 
of the axion density involves an integration of this spectrum over the range $n=n_1$ to $n=n_2$, 
where the frequencies $n_1\propto 1/t$ and $n_2\propto \fa$ correspond to the long distance and 
short distance cut-offs, naturally given by the radius of curvature of the string $(\sim t)$ and 
its width $(\sim\fa^{-1})$.
In the case where $q>1$ then the dependence on the high frequencies is proportional to $(n_2/n_1)^{q+1}$
(plus an ${\cal O}(1)$ factor dependent on $q$) and, therefore, in all realistic cosmological scenarios 
this dependence is weak and effectively can be ignored. However, the situation is very much different 
if $q=1$. In this case the integral is logarithmically divergent - hence, it is called a `flat spectrum' -
and the axion density is proportional to 
$\log(n_1/n_2)\sim 100$. At the simplest level, it is this factor 100 which is responsible for the two 
orders of magnitude discrepancy in predictions for the axion mass. Hence, from the point of view of 
axion cosmology the important question is which is the better approximation: (i) a spectrum dominated
by the lowest harmonics $q>1$ or (ii) the flat spectrum $q=1$.

In the rest of this article we shall present analytic and numerical arguments which clearly suggest 
that $q>1$. First, we will summarize the basic points of work presented in ref.~\cite{BatShe94a} which 
confirms the efficacy of the Kalb-Ramond (KR) action for describing the dynamics of global strings by 
comparison to numerical simulations. In doing so we will respond to various criticisms which have 
been made about our methods, showing them to be groundless. This work supports very strongly the 
basic physical premises of the original picture for global string radiation put forward by Davis. Then we will 
report on new work in which we evolve a small network of global strings~\cite{BatShe98}. Although these simulations 
are the largest ever done $(300^3$ points), they are still very small relative cosmological scales and 
the results must be correctly interpreted. Making sensible extrapolations to cosmological scales, they 
also confirm unequivocally the view that $q>1$.

\begin{figure}[t]
\centerline{\psfig{file=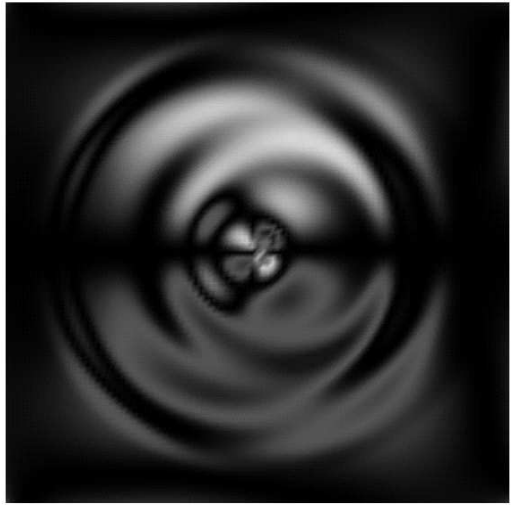,width=2.75in}}
\caption{Axion radiation from an oscillating periodic string configuration in a three-dimensional field theory simulation. In this perpendicular cross-section, the string oscillates horizontally. Note the dominance of the $n$=$2$ quadrupole radiation} 
\label{fig-rad}
\end{figure}

\section{RADIATION FROM OSCILLATING STRINGS}

The basic features of global cosmic strings can be described by the Goldstone model with a global U(1) symmetry for a complex scalar field $\Phi$, whose action is 
\begin{equation}
S=\int d^4x\left[\partial_{\mu}\bar\Phi\partial^{\mu}\Phi-{1\over 4}\lambda\left(\bar\Phi\Phi-\fa^2\right)^2\right]\,.
\label{gold-action}
\end{equation}
This can be transformed into the KR action for line-like strings coupled to an antisymmetric tensor field $B_{\mu\nu}$,
\[
S=-\mu_0\int\sqrt{-\gamma}d\sigma d\tau+{1\over 6}\int H^2 d^4x
\]
\begin{equation} 
~~~~~~~~~~~~~ -2\pi\fa\int B_{\mu\nu}d\sigma^{\mu\nu}\,,
\label{kr-action}
\end{equation}
by exploiting the duality between a massless scalar field and the antisymmetric tensor $B_{\mu\nu}$.
A detailed explanation of the derivation of this action and the notation used here can be found in ref.~\cite{BatShe97}. For the current discussion, all we need to appreciate is that single string can be represented within both formalisms, in the field theory case by a vortex solution to the field equations and in the KR action by a solution to the Nambu equations of motion modified to include the effects of radiation backreaction. If $X^{\mu}(\sigma,\tau)$ are the coordinates of the string, then the equation of motion is 
\begin{equation}
\mu_0(\ddot X^{\mu}-X^{\mu\prime\prime})=F^{\mu}_{\rm self}+F^{\mu}_{\rm rad}\,,
\label{nambu-eqn}
\end{equation}
where $F^{\mu}_{\rm self}$ and $F^{\mu}_{\rm rad}$ are the contributions to the backreaction from the self and radiation fields respectively. The dominant divergent part of the self force is given by $F^{\mu}_{\rm self}=-2\pi\fa^2\log(\Delta/\delta)(\ddot X^{\mu}-X^{\mu\prime\prime})$ leading to the well known renormalization of the mass per unit length
\begin{equation}
\mu(\Delta)=\mu_0+2\pi\fa^2\log(\Delta/\delta)\,.
\label{mass-renorm}
\end{equation}
This renormalization is critical to our argument based on the Kalb-Ramond action since, if we assume the effects of the radiation field are small, then the zeroth-order string solution is just that of the simple Nambu equations of motion and the effects of radiation can be thought of as being small perturbations. It is analogous to the mass renormalization of the point electron used in electromagnetism, and a number of papers in the literature support it~\cite{BatShe94c,DabQua90,CHH90,DamBou}.

To proceed analytically from this point we just need solutions to the Nambu equations of motion and then 
the power spectrum of axion radiation can be computed using a formula similar to the quadrupole formula 
used for gravitational radiation~\cite{VilVas87}. We investigated the spectrum emitted by a wide range of loop 
and periodic infinite string solutions, these included the Kibble-Turok loops - of which the circular loop 
is a special case, a kinky loop, a helicoidal long string, a sinusoidal long string and a kinky long string. 
(Note that `kinks' correspond to discontinuities in the tangents and velocities along the string, which 
propagate at the speed of light.)
In the only exactly tractable case, the helicoidal string, the power spectrum at large $n$ was shown to be 
exponential and various approximate arguments suggested that the quadrupole mode was the first for any 
exactly periodic solution. All the other solutions studied could be
shown asymptotically to produce the generic $q>1$ spectrum, but for two 
exceptions.  First, a perfectly circular loop because of its special 
symmetry collapses to a point and produces a divergent $q=1$ spectrum.  
However, unlike this special case, loops produced by a realistic network 
can be expected to have a spectrum at least $q\ge 4/3$.  Secondly, 
a periodic kinky long string solution can be written as the sum of an infinite series 
of sinusoidal perturbations, so this too has an apparently divergent spectrum with $P(n)\propto n^{-1}$ for large $n$.
However, as we discuss below, radiation backreaction rapidly damps out
the higher frequency modes, effectively turning the initial kink into a
sinusoidal perturbation radiating into the lowest modes.

A complementary numerical approach can be taken by evolving the field equations corresponding to the action 
(\ref{gold-action}) on a discretized grid ($100^3$ points). In order to compare directly with the analytic 
calculations discussed above, we concentrated on perturbed periodic long string configurations, in particular 
those with sinusoidal and kink perturbations. Separation of the self-field from that of the radiation poses a 
more serious problem in this approach than in the KR formalism. In fact there is no exact way of doing this, 
but it can be done approximately, the errors being largest close to the string core. The method we use is to 
wait until the string becomes straight and then subtract out the static solution. 
As we have already noted in our discussion of the KR formalism,
 the self-field of the string obeys the same equation of motion as the string itself allowing the renormalization of the 
string tension (\ref{kr-action}).  Hence, as the string moves, the
self-field moves with it, maintaining approximately the same form as the
static solution.  Obviously, Lorentz contractions 
of the string may distort the string, but this will only become
important at large amplitude when string velocities $v\approx c$.

Perhaps the most compelling evidence for the validity of the subtraction comes from the results 
themselves since they are so clean. As shown in Fig.~\ref{fig-rad} the radiation field for a
sinusoidal perturbation comprizes of almost perfect quadrupole lobes characteristic of the 
$n$=$2$ mode, which can be seen to propagate outwards from the centre of the string as the system 
evolves. It has been suggested that this travelling wave phenomena could be an artefact of the 
subtraction, but this is simply not credible! Further confirmation comes from spectral analysis 
of the radiation field. We find that the 
spectrum is dominated by the $n$=$2$ mode with a rapid fall-off into the higher modes; any problems 
with the subtraction would surely create high frequency noise. Further discussion of this issue will 
be included in ref.\cite{BatShe98b}.

If one sets up an initial kink solution, the spectrum of radiation initially contains more power in higher frequency modes, 
as suggested by the analytic calculations. But as the solution evolves higher frequency radiation appears 
to be damped out much faster than the lower frequencies, close 
to the characteristic frequency of the perturbation. Soon the spectrum of radiation becomes indistinguishable 
from that of the sinusoidal perturbation.  The initially 
sharp profile of the kink becomes visibly rounded, so that after around 5 oscillations the solution  
looks very much like 
a sinusoidal perturbation. This strongly suggests that high frequency radiation is damped much more quickly by radiation 
backreaction, generically leaving behind the dominant lower harmonics~\cite{BatShe94c}.

\begin{figure}[t]
\centerline{\psfig{file=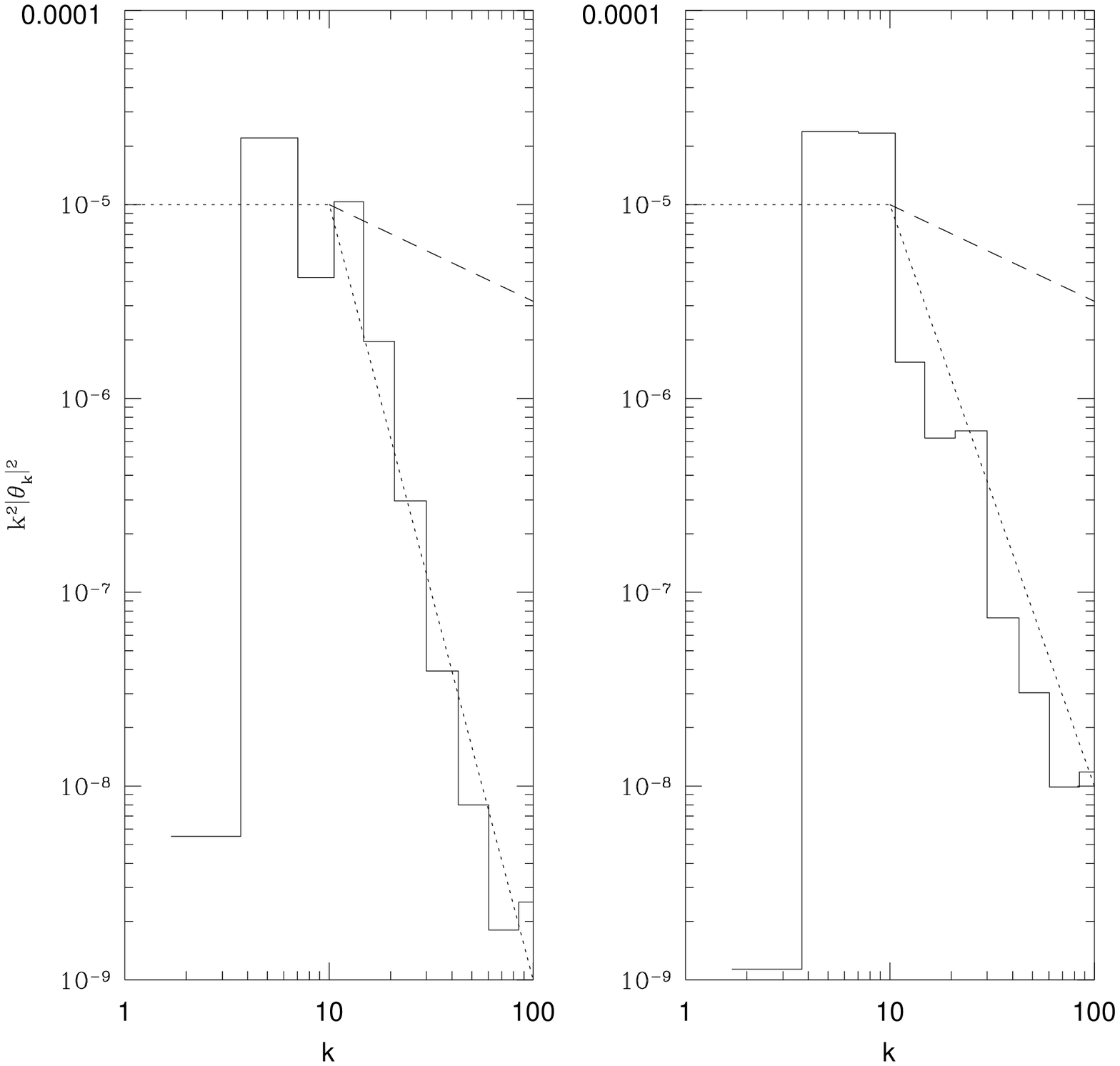,width=2.75in}}
\caption{The results of 3D spectral analysis of the argument of the complex scalar field in numerical simulations. We plot $|E_{k}|=k^2|\theta_k|^2$ against $k$, which is the square root of the spectrum of the axion energy density. On the right is that from the collapse of loops, included are lines representing $k^{-4}$ (dotted line) and $k^{-1/2}$ (dashed line), and on the left from the collapse of strings connected to domain walls, with lines for $k^{-3}$ (dotted line) and  $k^{-1/2}$ (dashed line).} 
\label{fig-spec}
\end{figure}

\section{THE COLLAPSE OF A NETWORK OF LOOPS}

In order to go further than the single string simulations discussed in the previous section, we have evolved a network of strings 
in a $300^3$ box with periodic boundary conditions. The intention here is not to investigate the scaling properties of the 
network, but rather to analyze the spectrum of radiation emitted by more general string configurations. The initial 
conditions were created by laying down a random noise configuration, before initially evolving it in a dissipative regime
(removing the radiation background). 
This process simulates some aspects of a cosmological phase transition. However, for our purposes the precise details 
of the initial distribution of strings are not important, the only salient point being that a simple random network 
of loops (due to the periodicity) can be created which in many ways mimics that formed in the early universe.

As the network evolves relativistically some of the loops collapse and emit massless axion radiation, while others 
interact with each other creating larger and smaller loops. This process continues until no more interactions 
are possible, with all loops collapsing into axions. The spectrum of loop sizes creating the axions has a fairly
wide random distribution, 
being a function of the initial conditions and the subsequent dynamics.  Once all the strings have collapsed 
all that is left is axion radiation, so there is now no self-field associated with the strings and hence we 
do not have to worry about any subtraction `ambiguities' before performing a 3D spectral analysis. We have 
analyzed the argument of the field $\Phi$, which corresponds to the massless axion field, ignoring the massive 
radiation emitted by strings on these small lengthscales.  This is plotted as $|E_k|=k^2|\theta_k|^2$ in 
Fig.~\ref{fig-spec}(a), which is the square root of the power spectrum of the axion contribution to 
the energy density. At small k - that is,  
large wavelengths - the spectrum appears to be dominated by the initial
correlation length of the strings, that is, wavelengths corresponding to
the most likely loops in the simulation (well above the string width). 
However, the focus of our attention is on the distribution of high 
frequency modes.  For $k>10$, we see that the spectrum of axions present
can be approximated by a fairly steep power law $|E_k|\propto k^{-4}$.  
Despite the considerable uncertainties in the exponent (which we shall
discuss elsewhere~\cite{BatShe98b}), these results are undoubtedly
inconsistent with the flat $q=1$ spectrum which should
fall-off here at the much slower rate $k^{-1/2}$.  

We have also performed simulations starting with the same initial conditions, but with the 
axion mass switched on from the start. In the original simulations the action was rescaled 
so that the problem is dimensionless and $\fa=1$ in these units. When the axion mass is 
included the value we use is $\ma=0.5$, which is very large compared with the cosmologically interesting value. 
But for the mass to have an effect in these simulations on a reasonable timescale one requires 
$\ma={\cal O}(\fa)$. This should be kept in  mind when interpreting these preliminary results. 

We find that the network collapses much more quickly once the mass is included since 
domain walls are formed connected to the strings. This hybrid network is unstable topologically, 
and the relativistic dynamics of the strings slices up the network creating small pieces of 
domain wall bounded by strings, which  subsequently decay into axions. Once all the pieces have 
collapsed we analyzed the spectrum of axions left in the box. As one can see from Fig.~\ref{fig-rad}(b) the 
spectrum can be approximated by a power law, this time something close to the $k^{-3}$, which is
again
definitely not consistent with the flat spectrum. Therefore, we have shown that in the QCD regime 
where the axions mass `switches on', the spectrum of axions emitted may have a slightly 
harder spectrum with more power in high frequency modes, albeit for small networks with 
an artifically high mass.  However, there appears to be no evidence whatever for anything even close to a flat spectrum.

We note, however, that it would be legitimate to criticise these
simulations on two important counts.  First, the size of the simulations is very small compared 
to the cosmological scales  we are trying to model. The renomalized mass per unit length is 
$\mu(\Delta)\approx 3\mu_0$, as compared to the realistic value of $\mu(\Delta)\approx 100\mu_0$, 
making the radiative decay of the strings much more rapid than one would expect for a cosmological network. 
However, such a point could only have some veracity if we were trying to defend a spectrum which 
was close to flat, since the damping in the simulations is much closer to `critical'. 
Secondly, the mass of the axion used in the domain wall simulations is far higher than
that for the early universe. However, the effect of an unphysically large mass should be to speed up 
the annihilation process, making higher frequency modes more prevalent than in a cosmological setting.
We shall return to these issues at greater length in a subsequent
publication~\cite{BatShe98b}.

\section{DISCUSSION}

In this paper we have attempted to elucidate the reason why the technical 
issue of the power spectrum of radiation emitted by global strings is important 
for axion cosmology and hence for the large-scale axion dark matter searches. 
At a first approximation for an order of magnitude estimate of the axion mass, 
all one has to do is decide between the two qualitatively different spectra with 
exponents $q>1$ and $q=1$. We have summarized a number of ways, both analytic and numerical, 
by which we can gain insight into what the true spectrum of axion string radiation actually is.
In each case, these approaches demonstrate a  $q>1$ spectrum dominated by the lowest
harmonics, one which is far removed from the flat $q=1$ spectrum  required for a low axion mass
$\ma \sim 1\mu$eV. 
While each of these approaches has caveats, the 
consistency of the picture they present is compelling!  
Having established the underlying axion spectrum on a firmer foundation, 
the time has come to unite our efforts in making accurate predictions
for the mass of a dark matter axion
by addressing the {\it real} quantitative issues that remain.


\def\jnl#1#2#3#4#5#6{\hang{#1, {\it #4\/} {\bf #5}, #6 (#2).}}


\def\jnlerr#1#2#3#4#5#6#7#8{\hang{#1, {\it #4\/} {\bf #5}, #6 (#2).
{Erratum:} {\it #4\/} {\bf #7}, #8.}}


\def\jnltwo#1#2#3#4#5#6#7#8#9{\hang{#1, {\it #4\/} {\bf #5}, #6 (#2);
{\it #7\/} {\bf #8}, #9.}}

\def\prep#1#2#3#4{\hang{#1 (#2),  #4.}}

\def\myprep#1#2#3#4{\hang{#1 (#2), '#3', #4.}}

\def\proc#1#2#3#4#5#6{\hang{#1 (#2), `#3', in {\it #4\/}, #5, eds.\ (#6).}
}
\def\procu#1#2#3#4#5#6{\hang{#1 (#2), in {\it #4\/}, #5, ed.\ (#6).}
}

\def\book#1#2#3#4{\hang{#1 (#2), {\it #3\/} (#4).}
									}

\def\genref#1#2#3{\hang{#1 (#2), #3}
									}


\def\prl{Phys.\ Rev.\ Lett.}
\def\pr{Phys.\ Rev.}
\def\pl{Phys.\ Lett.}
\def\np{Nucl.\ Phys.}
\def\prp{Phys.\ Rep.}
\def\rmp{Rev.\ Mod.\ Phys.}
\def\cmp{Comm.\ Math.\ Phys.}
\def\mpl{Mod.\ Phys.\ Lett.}
\def\apj{Ap.\ J.}
\def\apjl{Ap.\ J.\ Lett.}
\def\aap{Astron.\ Ap.}
\def\cqg{Class.\ Quant.\ Grav.} 
\def\grg{Gen.\ Rel.\ Grav.}
\def\mn{M.$\,$N.$\,$R.$\,$A.$\,$S.}
\def\ptp{Prog.\ Theor.\ Phys.}
\def\jetp{Sov.\ Phys.\ JETP}
\def\jetpl{JETP Lett.}
\def\jmp{J.\ Math.\ Phys.}
\def\cupress{Cambridge University Press}
\def\pup{Princeton University Press}
\def\wss{World Scientific, Singapore}


\begin{thebibliography}{9}
 
\bibitem{Dav86}
\jnl{R.L. Davis}{1986}{}{\pl}{180B}{225}

\bibitem{DavShe89}
\jnl{R.L. Davis and E.P.S. Shellard}{1989}{Do axions need inflation?}
{\np}{B324}{167}

\bibitem{BatShe94a}
\jnl{R.A. Battye and E.P.S. Shellard}{1994}{Global
string radiation}{\np}{B423}{260}

\bibitem{BatShe94b}
\jnlerr{R.A. Battye and E.P.S Shellard}{1994}{}{\prl}{73}{2954}{76}{2203}

\bibitem{BatShe97}
\genref{R.A. Battye and E.P.S Shellard}{1997}{Recent Perspectives on 
Axion Cosmology}{in Proceedings of {\it 
Dark Matter in Astro- and Particle Physics}, ed.\ H.V. Klapdor-Kleingrothaus
\& Y. Ramachers, pp. 554 (World Scientific, 1997), also astro-ph/9706014}

\bibitem{HarSik87}
\jnl{D.D. Harari and P. Sikivie}{1987}{On the evolution of global strings 
in the early universe}{\pl}{195B}{361}

\bibitem{HagSik91}
\jnl{S.H. Chang, C. Hagmann and P. Sikivie}{1991}{Computer simulations of the motion and
decay of global strings}{\np}{B363}{247}

\bibitem{BatShe94c}
\jnl{R.A. Battye and E.P.S. Shellard}{1994}{}{\prl}{75}{4354}
\jnl{R.A. Battye and E.P.S. Shellard}{1996}{}{\pr}{D53}{1811}

\bibitem{HagSik98}
\prep{C. Hagmann and P. Sikivie}{1998}{}{In these proceedings}

\bibitem{BatShe98}
\prep{E.P.S. Shellard and R.A. Battye}{1998}{}{In these proceedings}

\bibitem{BatShe98b}
\prep{R.A. Battye and E.P.S. Shellard}{1998}{}{In preparation}

\bibitem{DabQua90}
\jnl{A. Dabholkar and J.M. Quashnock}{1990}{Pinning down the
axion}{\np}{B333}{815}

\bibitem{CHH90}
\jnl{E. Copeland, D. Haws and M. Hindmarsh}{1990}{Classical theory of radiating strings}{\pr}{D42}{726} 

\bibitem{DamBou}
\prep{A. Buonanno and T. Damour}{1998}{Effective action and tension renormalisation
for cosmic and fundamental strings}
{hep-th/9803025}

\bibitem{VilVas87}
\jnl{A. Vilenkin and T. Vachaspati}{1987}{Radiation of Goldstone bosons from
cosmic strings}{\pr}{D35}{1138}

\end{thebibliography}
\end{document}